\documentclass[showpacs,prb,12pt,onecolumn]{revtex4}

\usepackage{graphicx}%
\usepackage{amssymb}
\begin{document}

%\preprint{HEP/123-qed}

\title{KAgF$_{3}$: quasi-one-dimensional magnetism in three-dimensional magnetic ion sublattice}% Force line breaks with \\
\author{Xiaoli Zhang, Guoren Zhang, Ting Jia, Ying Guo, and Zhi Zeng\footnote{Correspondence author. Email: zzeng@theory.issp.ac.cn}}
\affiliation{Key Laboratory of Materials Physics, Institute of Solid
State Physics, Chinese Academy of Sciences, Hefei 230031, P. R.
China}
\author{H. Q. Lin}
\affiliation{\it Department of Physics and Institute of Theoretical
Physics, The Chinese University of Hong Kong, Shatin, Hong Kong, P.
R. China}
\date{\today}

%\date{\today}% It is always \today, today, but you may specify any date with \date.
\begin{abstract}
The electronic structure and magnetic properties of the
Jahn-Teller-distorted perovskite KAgF$_{3}$ have been investigated
using the full-potential linerized augmented plane-wave method. It
is found that KAgF$_{3}$ exhibits significant quasi-one-dimensional
antiferromagnetism with the ratio of exchange constant
$|$\emph{J$_{\bot}$}$|$ (perpendicular to the \emph{z} axis) and
\emph{J} (along the \emph{z} axis) about 0.04, although the
sublattice of magnetic ion is three-dimensional. The strong
quasi-one-dimensional antiferromagnetism originates from the
\emph{C}-antiferro-distortive orbital ordering of the Ag$^{2+}$
4\emph{d}$^{9}$ ions. The orbital ordered antiferromagnetic
insulating state in KAgF$_{3}$ is determined by on-site Coulomb
repulsion to a large extent.
\end{abstract}

\pacs{75.25.-j, 71.20.-b}\maketitle
%\date{\today}% It is always \today, today, but you may specify any date with \date.
\newpage

\section{\bf Introduction}
Low-dimensional magnets have unique electronic and magnetic
properties. For example, the physical properties of
quasi-two-dimensional (2D) high-temperature
superconductor\cite{Shirane, Kamihara}, spin-Peierls
compound\cite{Shaz}, Haldane chain\cite{Tsunetsugu} and
spin-ladder\cite{Pieper} as well as spin-frustrated
systems\cite{Kageyama, Pandey} have been extensively investigated.
In most of the low-dimensional magnets, the magnetic cation
sublattices consist of planes or chains which are kept reasonably
far apart. Materials which exhibit three-dimensional (3D) magnetic
ion sublattice and low-dimensional magnetic behavior have been of
interest to both experimentalist and
theorists.\cite{Tong,Satija,Towler,Giovannetti,Xiao} Up to date, the
electronic orbital ordering (OO) is considered as an essential
factor in determining the strong spacial exchange anisotropy in
these materials.

KAgF$_{3}$ is a material which exhibit strong spacial exchange
anisotropy within 3D magnetic Ag$^{2+}$ ion sublattice.\cite{Mazej}
It was first synthesized in 1971 by Odenthal \emph{et
al}.\cite{Odenthal} and received, as well as other silver fluorides,
attention due to the pursuit of
superconductivity\cite{Grochala1,Grochala2,Grochala3} in
transition-metal compounds other than the cuprates. Previous
experimental studies\cite{Grochala3,Mazej} have provided the
evidence of its antiferromagnetism with T$_N$=64 K as well as the
insulating characteristic below this temperature. In contrast to the
undoped cuprates that are quasi-2D antiferromagnetic (AFM)
insulator, KAgF$_{3}$ was reported as a quasi-one-dimensional (1D)
antiferromagnet using LSDA+\emph{U} method.\cite{Mazej} However,
more efforts are still needed to well understand the origin of the
quasi-1D antiferromagnetism.

In the present work, we investigate the electronic structure and the
magnetic properties of KAgF$_{3}$ using density functional theory
(DFT). The calculated \emph{J$_{\bot}$/$\mid$J$\mid$}=0.04 confirms
that the compound is a quasi-1D antiferromagnet. Here, \emph{J}
(\emph{J$_{\bot}$}) refers to exchange constant along (perpendicular
to) the \emph{z} axis. The quasi-1D antiferromagnetism can be
understood by the \emph{C}-antiferro-distortive OO which can be
obtained in the presence of on-site Coulomb repulsion \emph{U}. The
picture of OO and AFM insulating state is thus very similar to that
of an isoelectronic and isostructural compound, KCuF$_{3}$.
\section{\bf CALCULATION METHODS}
Our electronic structure calculations are performed within the
full-potential linearized augmented plane-wave
framework.\cite{Blaha} The generalized gradient approximation (GGA)
of Perdew-Burke-Ernzerhof form is adopted.\cite{Perdew} To include
the on site Coulomb interaction GGA+\emph{U} (\emph{U}=on-site
Coulomb repulsion strength) approach is used with
\emph{U$_{eff}$}=\emph{U}-\emph{J} (\emph{J} is the exchange
interaction) instead of \emph{U}.\cite{Anisimov} Here in our work,
on-site Coulomb repulsion \emph{U} is applied to Ag 4\emph{d}
orbitals only. The Muffin-tin sphere radii are chosen to be 2.24,
2.09, and 1.85 bohr for K, Ag, and F atoms, respectively. Within the
Muffin-tin sphere the electrons behave as they were in the free
atom, and the wave functions are expanded using radial functions
(solutions to the radial part of Schr\"{o}inger equation) times
spherical harmonics. Out of the Muffin-tin spheres wave functions
are expanded using plane waves. The value of R$_{MT}$K$_{max}$ (the
smallest muffin-tin radius multiplied by the maximum \emph{k} value
in the expansion of plane waves in the basis set) is set to 7.0. We
use 500 \emph{k} points (i.e., 192 \emph{k} points in the
irreducible wedge of the Brillouin zone) for the integration over
the Brillouin zone. Self-consistency was considered to be achieved
when the total energy difference between succeeding iterations is
less than 10$^{-5}$ Ry/unit cell. The present setup ensures a
sufficient accuracy of the calculations.

In our study, we carry out total energy calculations with three
distinct spin states of (1) ferromagnetic (FM), (2) \emph{A}-AFM
(ferromagnetism in \emph{xy} plane, AFM stacking), and (3)
\emph{G}-AFM (antiferromagnetsm in \emph{xy} plane, AFM stacking).

\section{\bf RESULTS AND DISCUSSIONS}
The calculations were performed for orthorhombic structure proposed
by Mazej \emph{et al}.\cite{Mazej}, space group \emph{Pnma} (No.62),
with the lattice constants \emph{a}=6.2689 {\AA}, \emph{b}=8.3015
{\AA}, \emph{c}=6.1844 {\AA} (see Fig. 1). In our calculations we
use for each AgF$_6$ octahedron such corresponding local \emph{xyz}
coordinates that the \emph{z} axis is along the b axis and the
\emph{x} and \emph{y} axes are approximately along the inplane
nearest neighbor Ag-Ag directions. In KAgF$_{3}$, the Ag ion
sublattice forms a pseudocubic structure, in which Ag-Ag distance
along \emph{z} direction is slightly shorter than that in the
\emph{xy} plane. There are two types of F sites. One connects the Ag
ions along the \emph{z}-axis (named as F$_{ap}$) and the other
connects the Ag ions in the \emph{xy} plane (named as F$_{pl}$). The
AgF$_4$ planes are slightly puckered since the F$_{pl}$ ions have a
0.31 {\AA} displacement from the plane formed by Ag ions. Each Ag
and its six near F ions form a Jahn-Teller (JT) distorted AgF$_6$
octahedron. The distortion leads to the alternating long and short
Ag-F bonds along the \emph{x }and \emph{y} axes. The four shorter
Ag-F bonds are almost identical. The cooperative JT distortion
provides a signature of an orbital ordering (OO) at the Ag$^{2+}$
sites, which determines the electronic structure as well as the
magnetic properties.

In order to clarify the electronic structure and the origin of the
quasi-1D magnetism in KAgF$_3$, our calculations are designed in two
stages. We start the study within GGA. To account for a possible
lattice distortion, we carried out an optimization of the atomic
positions, keeping the unit-cell parameters fixed and relaxing the
atomic coordinates. The structural optimization shows that the
Ag-F$_{ap}$ bond is elongated by 0.012 {\AA} and the in-plane longer
(shorter) Ag-F$_{pl}$ bond is shortened (elongated) by 0.085 {\AA}
(0.038 {\AA}) as compared with the experimentally observed
structure. As a result, the large in-plane distortion of about 0.3
{\AA} found experimentally, decreases to 0.2 {\AA} theoretically.
Moreover, the GGA results for the optimized structure show that the
FM, \emph{A}-AFM and \emph{G}-AFM spin states are not stable and
converge to nonmagnetic (NM) metallic solution. Fig. 2 shows the
total and the orbital-resolved density of states (DOS) for the NM
KAgF$_3$. The distortive AgF$_6$ octahedron crystal field splits the
Ag$^{2+}$ 4\emph{d}$^{9}$ orbitals into fully occupied
\emph{t}$_{2g}$ (\emph{x}\emph{z}, \emph{x}\emph{y},
\emph{y}\emph{z}) and three fourth occupied \emph{e}$_g$
(\emph{z}$^{2}$-\emph{y}$^{2}$, 3\emph{x}$^{2}$-\emph{r}$^{2}$)
orbitals. Both the anti-bonding Ag \emph{d}$_{z^{2}-y^{2}}$ and
\emph{d}$_{3x^{2}-r^{2}}$ bands cross the Fermi level with large
band widths of about 3 eV and have strong covalency\cite{Grochala2}
with the F orbitals. The band center of the \emph{d}$_{z^{2}-x^{2}}$
is higher in energy than that of \emph{d}$_{3y^{2}-r^{2}}$ orbital.
Thus, the JT distortion generates a crystal-field splitting which is
not large enough to separate the two \emph{e}$_{g}$ bands. Even the
GGA calculations for the experimental structure still give a NM
metallic solution for all the enforced spin states, in accordance
with the previously published work.\cite{Mazej} Unlike the
electronic structure of GGA optimized structure with two type of
\emph{e}$_g$ bands crossing the Fermi level, the distortion alone
separates the \emph{e}$_g$ bands with only one of them cross the
Fermi level (not shown). Therefore, the GGA calculations can not
reproduce the insulating AFM nature of KAgF$_3$. This is in contrast
to another fuoroargentate Cs$_2$AgF$_4$ in which JT distortion alone
can lead to the orbital ordered insulating state.\cite{Wu,
Kasinathan,Hao}

As the second stage, we concern the correlation interaction by
GGA+\emph{U} approach. In general, strong on-site Coulomb repulsion
\emph{U} increases the localization of \emph{d} electrons and favors
orbital ordering and/or magnetic order. However, the magnitude of
the correlation for the Ag$^{2+}$ ion is not known. In order to
determine its value, we performed a structural optimization as a
function of \emph{U} (1-6 eV) and found the \emph{U} value with
which the equilibrium structure matches the experimental one. This
scheme to extract the Coulomb parameter is feasible because the
on-site Coulomb repulsion \emph{U} determines the orbital
polarization of the \emph{e}$_{g}$ states for a large extent, which,
in turn, causes the structural JT lattice distortion. The optimized
inequivalent Ag-F distances \emph{d}$_1$, \emph{d}$_2$ and
\emph{d}$_3$ are shown in Fig. 3. We find that the optimized
structure for \emph{U}=5.0 eV is the closest to the one obtained
experimentally. And thus this \emph{U} is adopted as the most
reliable one. This value is smaller than the commonly used
\emph{U}=7-9 eV value for Cu$^{2+}$ but still reasonable, because
the 4\emph{d} orbitals of an Ag$^{2+}$ ion are less contracted than
the 3\emph{d} orbitals of a Cu$^{2+}$ ion.

The GGA+\emph{U} (5 eV) total energy calculations of various spin
states shown in table I indicate that the ground spin state is the
\emph{A}-AFM insulating state. The obtained ground spin state is in
accordance with the previous LSDA+U results.\cite{Mazej} This
indicates that the on-site Coulomb repulsion is important in
determining the AFM insulating properties in KAgF$_3$. We note that
the difference in total energy between the \emph{A}-AFM and
\emph{G}-AFM spin states is as small as 23 meV/4f.u., while the
energy difference between the \emph{A}-AFM and FM spin states is
about 272 meV/4f.u.. We should notice the fact that \emph{A}-AFM and
\emph{G}-AFM spin states have opposite ordering in the
\emph{x}\emph{y} plane but the same ordering along the \emph{z}
direction, whereas both the \emph{A}-AFM and FM spin states have the
same ordering in the \emph{x}\emph{y} plane but different ordering
along the \emph{z} axis. This reveals that AFM exchange interactions
along the \emph{z} direction are more favorable and robust than FM
coupling in the \emph{x}\emph{y} plane, indicating a strong spacial
exchange anisotropy in KAgF$_3$.

To gain additional insight into the spacial exchange anisotropy, we
evaluate the magnetic exchange constants along and perpendicular to
the \emph{z} axis numerically in terms of the Heisenberg spin
Hamiltonian:
\begin{equation}
H=J\sum_{i,j}\textbf{S$_{i}$} \cdot \textbf{S$_{j}$} +
J_{\bot}\sum_{k,l}\textbf{S$_{k}$} \cdot \textbf{S$_{l}$}
\end{equation}
where the first (second) term refers to summing over all nearest
neighbors along (perpendicular to) the \emph{z} axis. By mapping the
obtained total energies for each magnetic state to the next nearest
Heisenberg model, the exchange interactions \emph{J} and
\emph{J$_{\bot}$} are
\begin{equation}
J=\frac{1}{8S^{2}}(E_{FM}-E_{\emph{A}-AFM})
\end{equation}
\begin{equation}
J_{\bot}=\frac{1}{16S^{2}}(E_{\emph{A}-AFM}-E_{\emph{G}-AFM})
\end{equation}
With the spin \emph{S}=1/2 of a 4\emph{d}$^{9}$ Ag$^{2+}$ ion, we
get \emph{J}=136 meV and \emph{J}$_{\bot}$=-5.8 meV, indicating
strong AFM coupling along the \emph{z} axis and a much weaker FM
coupling in the \emph{x}\emph{y} plane. \emph{$|J_{\bot}|$/J}=0.04
reflects strong quasi-1D magnetism in KAgF$_{3}$. Thus, it has been
concluded that KAgF$_{3}$ is another material exhibiting quasi-1D
magnetic behavior within 3D magnetic ions sublattice, which is in
accordance with the results of Mazej et al..\cite{Mazej} The
relatively large \emph{J} and \emph{J}$_{\bot}$ values are
consistent with the findings that the DFT electronic structure
calculations generally overestimate the magnitude of spin exchange
interactions.\cite{Whangbo}

The quasi-1D magnetic behavior can be understood from the ground
state electronic structure and OO of KAgF$_3$. The total and the
orbital-resolved DOS for the \emph{A}-AFM spin state experimental
structure are shown in Fig. 4. The Coulomb repulsion \emph{U} pushes
the occupied and unoccupied 4\emph{d} levels downward and upward
respectively and opens up an insulating gap. It is obvious that the
one hole states mainly occupy the higher level
\emph{d}$_{z^{2}-y^{2}}$ bands and the orbital polarization of
\emph{e}$_g$ states is enhanced. Actually, due to the strong Ag-F
covalency\cite{Grochala2} the one hole spreads over the six fluorine
atoms of the AgF$_6$ octahedron. This is also reflected by the
magnitude of local spin magnetic moments within each muffin-tin
sphere, $\pm$0.62 $\mu$$_{B}$/Ag, 0$\mu$$_{B}$/F$_{ap}$ and
$\pm$0.09$\mu$$_{B}$/F$_{pl}$ (see table I). The hole states have
alternating \emph{d}$_{z^{2}-y^{2}}$ and \emph{d}$_{z^{2}-x^{2}}$
symmetry for Ag$^{2+}$ ions in the xy plane (see the last panel in
Fig. 5) because of the cooperative JT distortion. The in-plane
\emph{d}$_{z^{2}-x^{2}}$/\emph{d}$_{z^{2}-y^{2}}$ OO repeats along
the \emph{z}-axis and \emph{C}-antiferro-distortive OO is formed,
which is clearly shown in Fig. 5. The overlap of Ag-F-Ag along the
\emph{z}-axis (see the upper and middle panel in Fig. 5) is larger
than that in the xy plane (the lower panel in Fig. 5) leading to the
dominate magnetic interaction along \emph{z}-axis. The
\emph{C}-antiferro-distortive OO would immediately give \emph{A}-AFM
spin state, according to Goodenough-Kanamori-Anderson rules. Also,
our GGA+\emph{U} calculation results confirm this mechanism since it
shows that the \emph{A}-AFM state is indeed more stable than the
\emph{G}-AFM and FM states (see Table I). Therefore, the results of
our calculations have clearly illustrated that KAgF$_3$ is an
orbitally ordered quasi-1D AFM insulator, in close analogy to
KCuF$_3$.

\section{\bf CONCLUSIONS }
By GGA and GGA+\emph{U} electronic structure calculations, we have
found the quasi-one-dimensional (1D) antiferromagnetism of KAgF$_3$
with \emph{J$_{\bot}$/$\mid$J$\mid$}=0.04, which may stimulate
further experimental studies. Here, \emph{J} (\emph{J$_{\bot}$})
refers to exchange constant along (perpendicular to) the \emph{z}
axis. The quasi-1D antiferromagnetism can be understood by the
\emph{C}-antiferro-distortive orbital ordering. The orbital ordered
insulating state can only be obtained when the on-site Coulomb
repulsion of Ag 4\emph{d} electrons is in consideration. With
\emph{U}=5 eV for Ag 4\emph{d} electrons the optimized structure is
in the best agreement with experiment. The picture of orbital
ordering and AFM insulating state is thus very similar to that of an
isoelectronic and isostructural compound, KCuF$_{3}$.

\section{\bf ACKNOWLEDGMENTS}
This work was supported by the special Funds for Major State Basic
Research Project of China(973) under Grant No. 2007CB925004,
Knowledge Innovation Program of Chinese Academy of Sciences under
Grant No. KJCX2-YW-W07, Director Grants of CASHIPS, and CUHK Direct
Grant No. 2160345. Part of the calculations were performed in Center
for Computational Science of CASHIPS.
%\bibliography{ref}

\clearpage

\begin{figure}[tp]
\vglue 1.0cm
\newpage

%---------------------------------Fig. 1---------------

\caption {(color online) Crystal structure of KAgF$_3$. The blue,
gray and yellow spheres represent K, Ag and F atoms, respectively.}

%---------------------------------Fig. 2---------------

\caption {(color online). The total and partial DOS of GGA optimized
structure for nonmagnetic state obtained in GGA. The Fermi level is
set at zero energy.}

%---------------------------------Fig. 3---------------
\caption {(color online). Deviation of relaxed inequivalent Ag-F
distances \emph{d}$_1$, \emph{d}$_2$ and \emph{d}$_3$ from the
experiment as a function of the Coulomb repulsion \emph{U}. The data
of \emph{U}=0 denotes the GGA optimized results.}

%---------------------------------Fig. 4-----------------
\caption {(color online). The total and partial DOS of KAgF$_{3}$
for \emph{A}-AFM spin state from GGA+\emph{U} with \emph{U}=5 eV.
The up and down panels in each DOS plot denote the spin-up and
spin-down states respectively.}

%---------------------------------Fig. 5-----------------
\caption {(color online). The contour plot of spin density in
\emph{xz}(upper panel), \emph{yz} (middle panel) and \emph{xy}
(lower panel) planes of KAgF$_{3}$ for \emph{A}-AFM state from
GGA+\emph{U} (5 eV). The solid (dot) lines depict the spin-up (
spin-down) states.}

\end{figure}

\begin{table}[tb]
%============================== Table 1========================
\caption{Electronic structure of KAgF$_{3}$ in FM, \emph{A}-AFM and
\emph{G}-AFM spin states obtained by GGA+\emph{U} with \emph{U}=5
eV. The total energy difference ($\Delta$\emph{E}, meV/4f.u.), the
energy gap (\emph{E}$_g$, eV), the local spin moment (\emph{SM},
$\mu$$_{B}$) of each Ag, apical F (F$_{ap}$), and planar F
(F$_{pl}$) atoms are shown.}

\begin{ruledtabular}
\begin{tabular}{ccccccccc}

&&$\Delta$$\emph{E}$&$$\emph{E}$_{g}$&Ag-\emph{SM}&F$_{ap}$-\emph{SM}&F$_{pl}$-\emph{SM}&&\\
 \hline
&FM &272&0.5&0.67&0.12 &0.1 \\
&\emph{A}-AFM &0&1.5&$\pm$0.62&0&$\pm0.09$ \\
&\emph{G}-AFM &23&1.4&$\pm$0.6&0&$\pm0.09$
\\
\end{tabular}
\end{ruledtabular}
\end{table}
\clearpage
%==========================  end Table 1===========================

%%%%%%%%%%%%%%%%%%%%%%The following is the figures%%%%%%%%%%%%%%%%%%%%%%%%%%%%%%%%%%%%%%%%%%%%%%%%%%%%%%%%%%%%
%---------------------------------------Fig. 1----------------------------------------------------
\begin{figure*}[htbp]
\center {$\Huge\textbf{Fig. 1  \underline{Zhang}.eps}$}
\includegraphics[bb = 10 10 1000 500, width=1.0\textwidth]{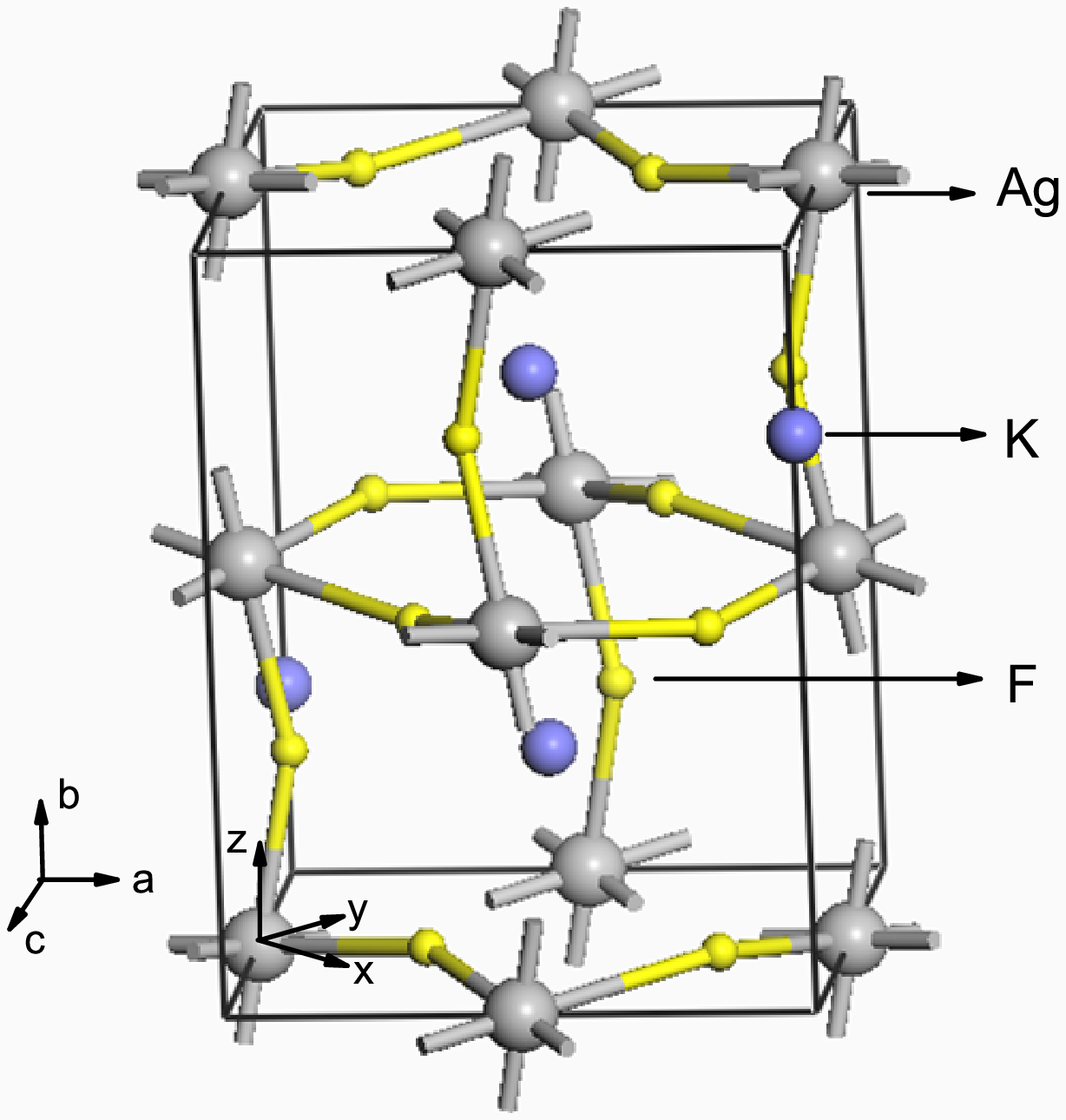}
\end{figure*}
%---------------------------------------Fig. 1----------------------------------------------------
\clearpage
\newpage

%---------------------------------------Fig. 2----------------------------------------------------
\begin{figure*}[htbp]
\center {$\Huge\textbf{Fig. 2  \underline{Zhang}.eps}$}
\includegraphics[bb = 5 5 1000 500, width=1.8\textwidth]{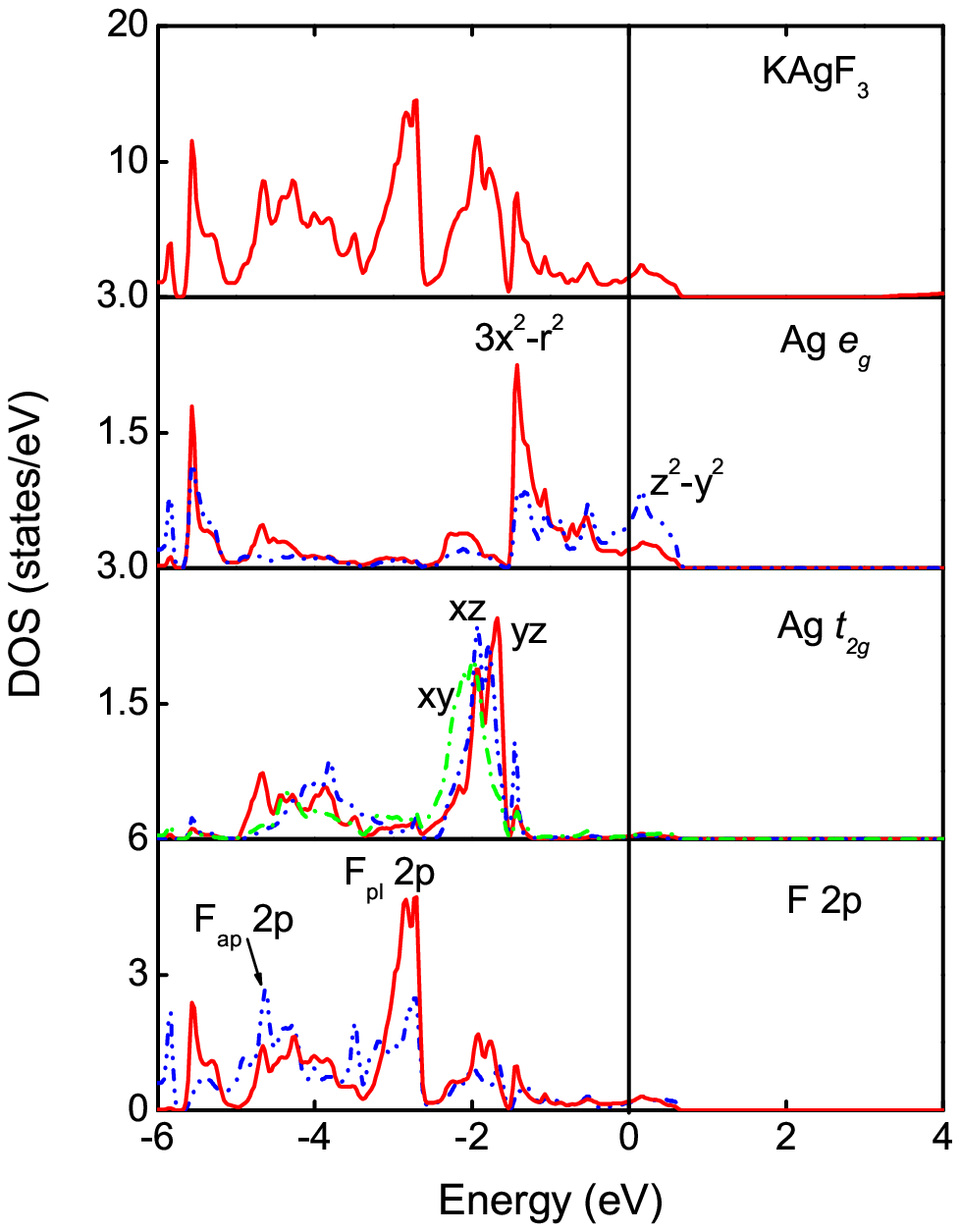}
\end{figure*}
%---------------------------------------Fig. 2----------------------------------------------------
\clearpage
\newpage

%---------------------------------------Fig. 3----------------------------------------------------
\begin{figure*}[htbp]
\center {$\Huge\textbf{Fig. 3  \underline{Zhang}.eps}$}
\includegraphics[bb = 10 10 1000 500, width=1.8\textwidth]{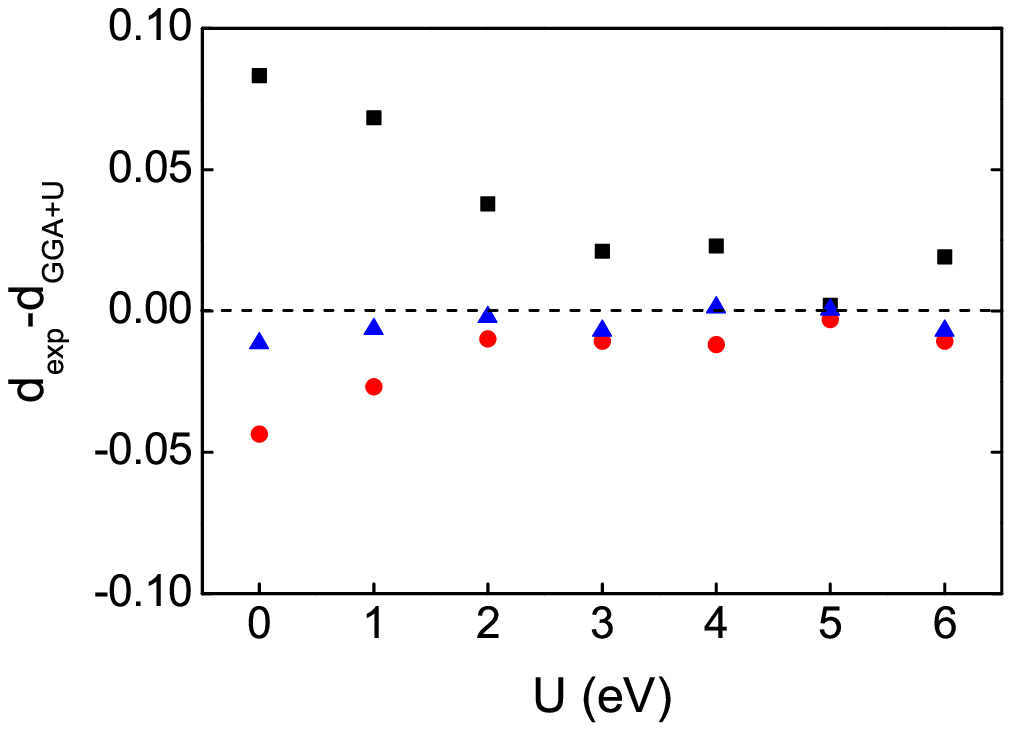}
\end{figure*}
%---------------------------------------Fig. 3----------------------------------------------------
\clearpage
\newpage
%---------------------------------------Fig. 4----------------------------------------------------
\begin{figure*}[htbp]
\center {$\Huge\textbf{Fig. 4  \underline{Zhang}.eps}$}
\includegraphics[bb = 10 10 1000 500, width=1.8\textwidth]{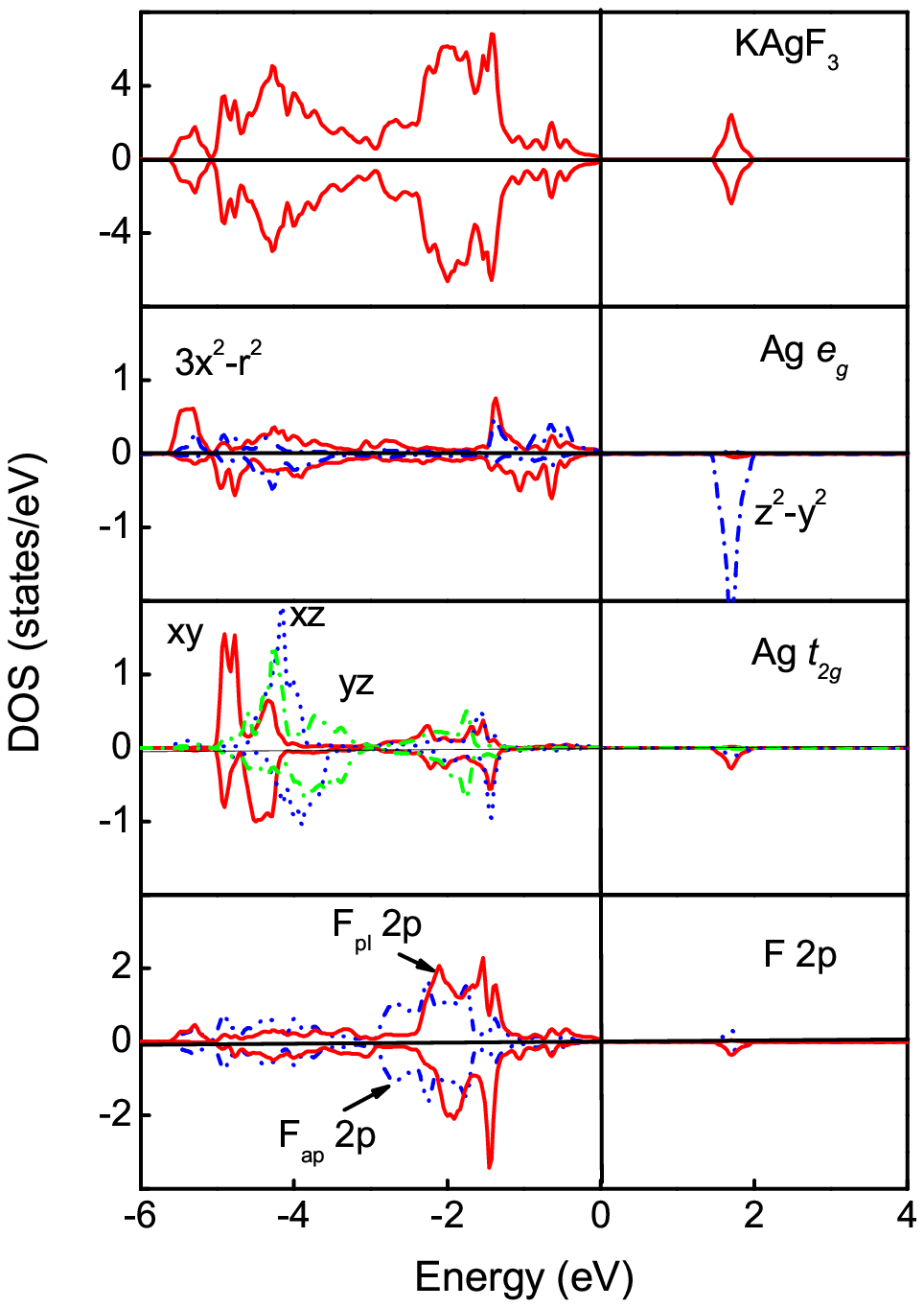}
\end{figure*}
%---------------------------------------Fig. 4----------------------------------------------------
\clearpage
\newpage

%---------------------------------------Fig. 5----------------------------------------------------
\begin{figure*}[htbp]
\center {$\Huge\textbf{Fig. 5  \underline{Zhang}.eps}$}
\includegraphics[bb = 10 10 1000 700, width=1.8\textwidth]{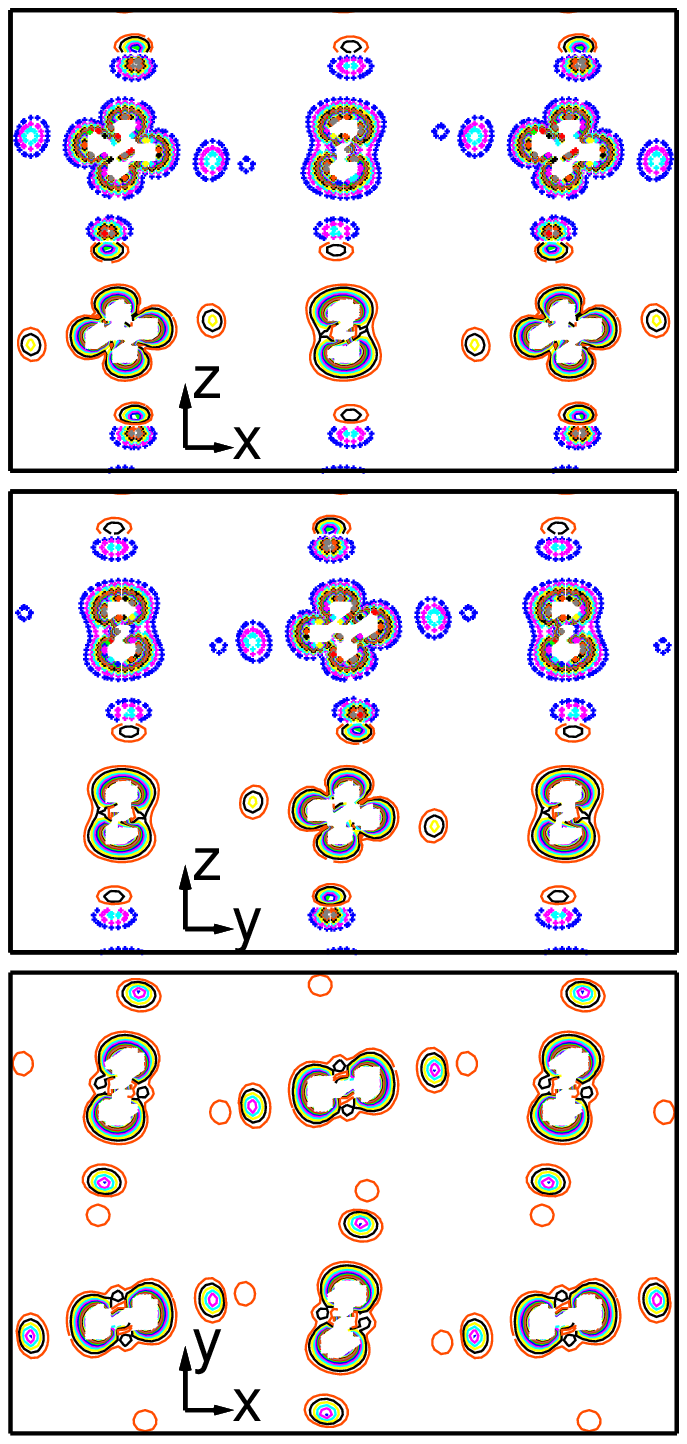}
\end{figure*}
%---------------------------------------Fig. 5----------------------------------------------------
\clearpage
\newpage

\end{document}